\documentclass{llncs}
\usepackage{mathptmx}       
\usepackage{helvet}         
\usepackage{courier}        
\usepackage{type1cm}        
\usepackage{makeidx}         
\usepackage{graphicx}        
\usepackage{multicol}        
\usepackage[bottom]{footmisc}
\usepackage{subfigure}
\usepackage{amsfonts}
\usepackage[cmex10]{amsmath}
\usepackage{color}

\begin{document}

\title{A Short Introduction to Local Graph Clustering Methods and Software}
\titlerunning{Hamiltonian Mechanics}  
%
\author{%
    Kimon Fountoulakis\inst{1} 
    \and 
    David F. Gleich\inst{2}
    \and
	Michael W. Mahoney\inst{3}
    }
\authorrunning{Fountoulakis et al.} 
%
\tocauthor{Kimon Fountoulakis and David F. Gleich and Michael Mahoney}
\institute{%
    University of Waterloo, Dept. of Computer Science, Waterloo ON N2L 3G1, Canada\\
	\and
	Purdue University, Dept. of Computer Science, West Lafayette, IN, USA\\
	\and 
	University of California Berkeley, ICSI and Dept. of Statistics, Berkeley, CA, USA
    }

\maketitle

Graph clustering has many important applications in computing, but due to the increasing sizes of graphs, even traditionally fast clustering methods can be computationally expensive for real-world graphs of interest.
Scalability problems led to the development of local graph clustering algorithms that come with a variety of theoretical guarantees~\cite{ACL2006}.
Rather than return a global clustering of the entire graph, local clustering algorithms return a single cluster around a given seed node or set of seed nodes. 
These algorithms improve scalability because they use time and memory resources that depend only on the size of the cluster returned, instead of the size of the input graph.
Indeed, for many of them, their running time grows linearly with the size of the output.

In addition to scalability arguments, local graph clustering algorithms have proven to be very useful for identifying and interpreting small-scale and meso-scale structure in large-scale graphs~\cite{LLDM2009,JBPMM2015}.
As opposed to heuristic operational procedures, this class of algorithms comes with strong algorithmic and statistical theory. These include statistical guarantees that prove they have \emph{implicit} regularization properties \cite{FKSCM2017,GM2014}.

One of the challenges with the existing literature on these approaches is that they are published in a wide variety of areas, including theoretical computer science, statistics, data science, and mathematics. 
This has made it difficult to relate the various algorithms and ideas together into a cohesive whole. 
We have recently been working on unifying these diverse perspectives through the lens of optimization~\cite{FDM2017} as well as providing software to perform these computations in a cohesive fashion~\cite{LGC2018}. 
In this note, we provide a brief introduction to local graph clustering, we provide some representative examples of our perspective, and we introduce our software named Local Graph Clustering (LGC).

%
\vspace{-\baselineskip}
\subsection*{Local graph clustering}

\vspace{-2mm}

Given a seed node, or a seed set of nodes, the goal of local graph clustering is to compute a cluster ``nearby'' the seed that is related to the ``best'' cluster nearby the seed. 
Here, ``best'' and ``nearby'' are intentionally left under-specified, as they can be formalized in one of a few different but related ways. 
For example, ``best''  is usually related to a clustering score such as conductance.  
Formally, local graph clustering can be easily understood as a recovery problem.
One assumes that there exists a target cluster in a given graph and the objective is to recover it from one or more example vertices inside the set. 
We can be more precise for a formulation involving conductance.  Assume that there exists a target cluster $B$ with conductance $\phi_T$ and we have one seed node in $B$, our objective is to find a cluster $A$ that resembles $B$ with conductance bounded by some function of $\phi_T$, where resemblance is captured here by normalized precision and recall. 
Moreover, we want to do this in running time and memory proportional to the size of $A$.

\begin{figure}[t]
\begin{minipage}{0.45\linewidth}
{\centering \includegraphics[width=0.7\linewidth]{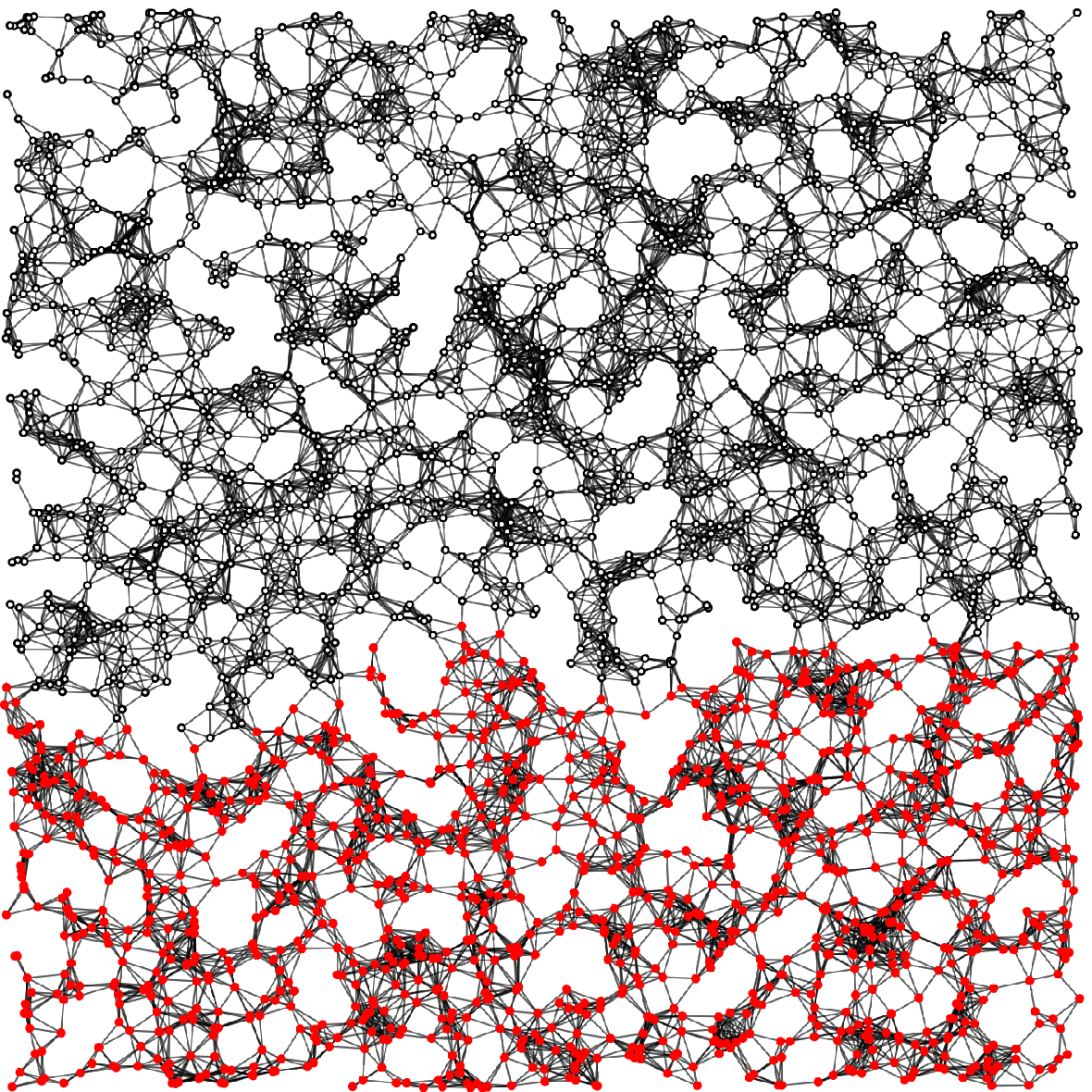}} \\
The near-optimal conductance solution for the random geometric graph bisects the graph into two large well-balanced pieces.
\end{minipage}\hfill 
\begin{minipage}{0.45\linewidth}
{\centering {\includegraphics[width=0.7\linewidth]{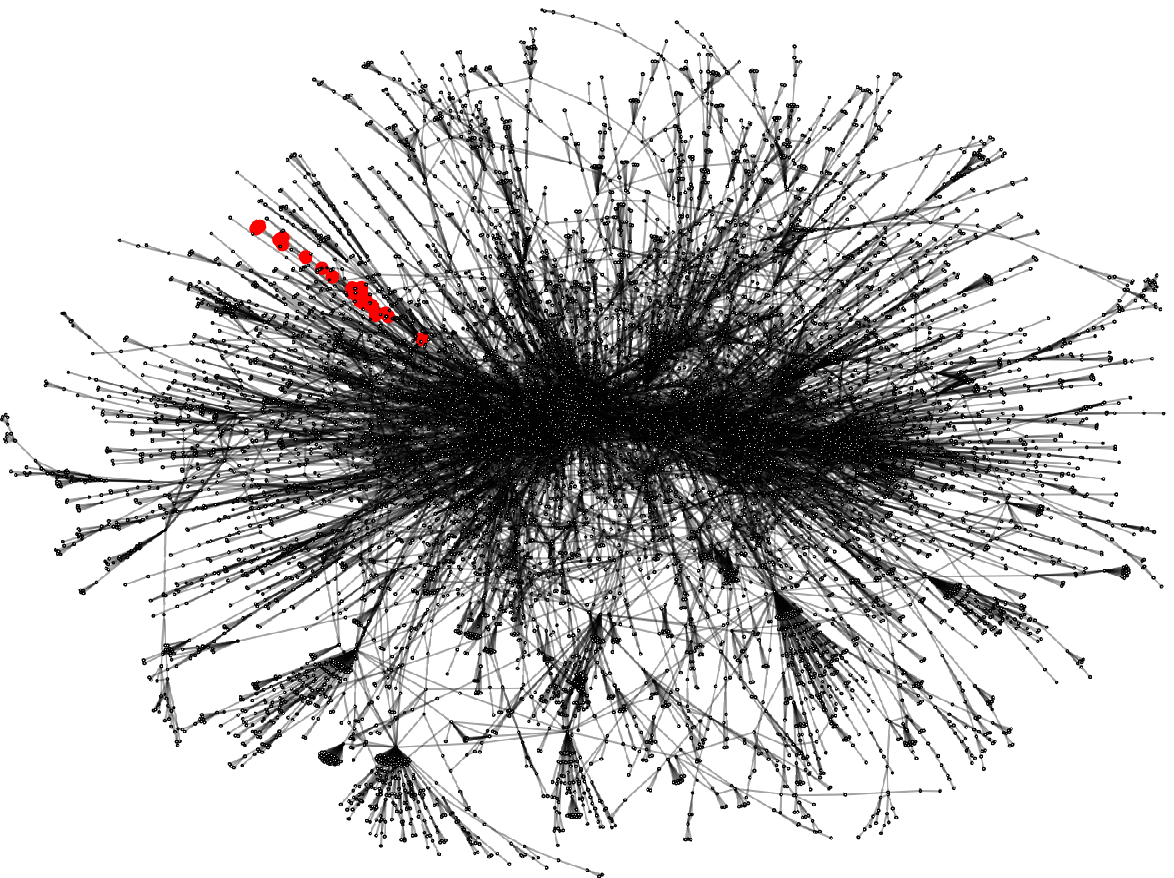}}\llap{\colorbox{white}{\includegraphics[scale=0.32]{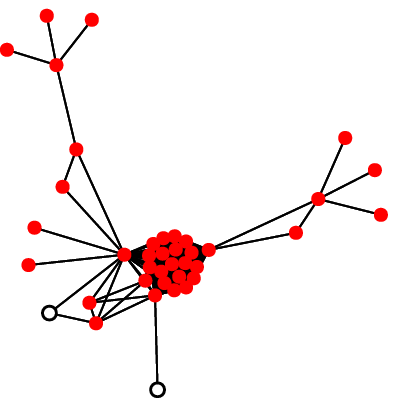}}} } \\
The near-optimal conductance solution for a typical data graph. (Inset. A zoomed view of the subgraph where two unfilled nodes are the border with the rest of the graph.)
\end{minipage}
	\caption{(Adapted from~\cite{FDM2017}.) 
    A motivation for local graph clustering. 
    In graphs with an underyling geometry (left), a good partition is a near bisection.
    In most data graphs (right)~\cite{LLDM2009,BPSDGA2004}, the good clusters and communities are small.  
    It is computationally ineffective to find the data-graph clusters using techniques whose runtime depends on the entire graph. 
    }
	\label{fig:small-cluster}
\end{figure}

As a quick example of why local graph analysis is frequently useful in data science applications, we present in Figure \ref{fig:small-cluster} the results of finding a good partition of both a random geometric graph and a more typical graph from machine learning and data science. 
In the random geometric graph, the size of best cluster or community is about half the graph. 
In this case, an algorithm with runtime that scales with the size of the graph is reasonable. 
In the graph that is more typical of machine learning and data science applications, the best cluster or community derived from a conductance metric is an exceedingly small fraction of the network. 
This means that standard graph algorithms whose runtime depends on the size of the  graph will do an enormous amount of work to return a tiny portion of the graph. 
Many other examples of this general phenomenon can be found~\cite{LLDM2009,JBPMM2015,LLDM2008,LLM2010}.
Local graph clustering techniques can find this cluster (and other similar clusters that happen not to be globally optimal) while touching not many more edges and nodes than are in the output cluster, greatly reducing the computation~time.


%
\vspace{-\baselineskip}
\subsection*{Our Software}

\vspace{-1mm}

Local Graph Clustering (LGC) is a Python package that uses C++ routines and brings scalable graph analytics on your laptop.
In particular, LGC provides methods that find local clusters, methods that improve a given cluster, tools to compute network community profiles, and multi-class label prediction.
The software is on GitHub \cite{LGC2018}.

\textbf{Methods in LGC.}\ LGC implements seven local graph clustering methods.\ 
Three spectral methods, i.e., approximate PageRank \cite{ACL2006}, PageRank Nibble \cite{ACL2006}, $\ell_1$-regularized PageRank \cite{FKSCM2017}, and four flow methods, i.e.,
Max-flow Quotient-cut Improvement \cite{LR2004}, FlowImprove \cite{AL2008}, SimpleLocal \cite{VGM2016} and Capacity Releasing Diffusion \cite{WFHM2017}.\

\textbf{Pipelines.}\ In LGC one can find pipelines which employ the above methods to compute network community profiles (NCPs) \cite{LLDM2008}.\
An NCP is a plot that is defined as the the quality of the ``best" community as a function of community size in a network.\ 
To measure the quality of a community we use conductance.\
Computing the NCP of a graph is an NP-hard problem, and therefore we compute an approximate version of it using local graph clustering methods.\
The same approximation has been suggested also in \cite{LLDM2009,LLDM2008}.\
LGC also implements multi-class label prediction using local graph clustering as a workhorse~\cite{GM2015}.\

\textbf{Scalability.} LGC offers routines to work with graphs that scale to the available memory of your system. We have used these routines to study graphs with billions of nodes on large memory machines, and graphs with $117$ million edges on a laptop computer by using about $9.4$ GB RAM.
Examples can be found in the GitHub repository~\cite{LGC2018}.\


\noindent
{\small
\textbf{Acknowledgements.}
This material is based on research sponsored by DARPA and the Air Force Research Laboratory (AFRL) under agreement number FA8750-17-2-0122, as well as the Army Research Office (ARO). The U.S. Government is authorized to reproduce and distribute reprints for Governmental purposes notwithstanding any copyright notation thereon. The views and conclusions contained herein are those of the authors and should not be interpreted as necessarily representing the official policies or endorsements, either expressed or implied, of DARPA and the AFRL or the U.S. Government.
}

\vspace{-\baselineskip}

\end{document}